# Critical current density and flux pinning in $La_{2-x}Pr_xCa_{2x}Ba_2Cu_{4+2x}O_z$ (x = 0.1 – 0.5) superconductors

S. Rayaprol, C. M. Thaker, N. A. Shah[+] and D. G. Kuberkar

*Department of Physics, Saurashtra University, Rajkot – 360 005 (India)*
[+]*Department of Electronics, Saurashtra University, Rajkot – 360 005 (India)*

**Abstract**

Polycrystalline $La_{2-x}Pr_xCa_{2x}Ba_2Cu_{4+2x}O_z$ (LPCaBCO) compounds with x = 0.1 – 0.5 were synthesized by solid-state reaction method and studied by room temperature X-ray diffraction, dc resistivity, dc magnetization and iodometry. The superconducting transition temperatures in these tetragonal triple perovskite compounds increases from 32 to 62 K ($T_c^{onset}$ values) with increasing dopant concentration. The mixing of rare earth $La^{3+}$ and $Pr^{3+/4+}$ ions at rare earth site ($La^{3+}$) along with substitution of divalent $Ca^{2+}$ results in the shrinkage of unit cell volume. The contraction of unit cell volume due to larger ion being substituted by smaller ions, gives rise to creation of pinning centers in the unit cell leading to increase in critical current density and flux pinning.

**Introduction**

Increasing the flux pinning energies has been one of the interesting challenges of high temperature superconductors both for understanding their basic phenomenon as well as for practical applications. Among various high temperature superconductors, RE-123 (RE = rare earth) type superconductors are favorite candidates for applications because of their well-developed pinning properties in magnetic field at low temperatures. From applications point of view, increase in current carrying capacity is always desirable and this can be achieved by effective flux pinning. This objective is achieved by the addition of normal state impurity phases (like R-211) [1], columnar defects generated by heavy ion irradiation [2], point defects generated by oxygen deficiency [3] or chemical substitution [3-4]. In this paper, we report the effect of simultaneous substitution of Pr and Ca in $La_2Ba_2Cu_4O_z$ *non-superconductor* on flux pinning ($F_p$) and critical current density ($J_c$) due to the creation of lattice disorder in the system.

The structural studies on $La_{2-x}Pr_xCa_{2x}Ba_2Cu_{4+2x}O_z$ (LPCaBCO) compounds with x = 0.1 – 0.5, have shown that these compounds exhibit stable tetragonal structure throughout the doping range [5]. It has been shown earlier that LPCaBCO and related RE (rare earth) substituted La-2125 type compounds can be derived from La-123 tetragonal family [6-8]. The interest in these compounds arises due to the increase in $T_c$ with increasing RE and Ca concentration. Superconductivity in La-2125 type superconductors appears due to the additional holes created by increasing Ca content. Like in RE-123 superconductors (except for Pr, Tb and Ce), RE *does not* directly affect the superconducting properties, as

Correspond Author: Dr. S. Rayaprol (TIFR, Mumbai) sudhindra@tifr.res.in,
Phone Number: +91-22-2280 4545 (Ext: 2439)

superconductivity is believed to originate in Cu-O planes.

In RE-123 systems, substitution of Pr at RE site results in suppression of superconductivity which has been attributed to either hole filling by mixed valent $Pr^{3+/4+}$ or hybridization between Pr (4f) and $CuO_2$ (2p) orbitals resulting in localization of mobile charge carriers [9, 10]. Doping divalent ions like $Ca^{2+}$ or $Sr^{2+}$ has revived superconductivity in these systems [11-14]. The substitution of Pr and Ca in LPCaBCO system thus presents an interesting case of simultaneous hole filling ($Pr^{3+/4+}$ on $La^{3+}$ site) and hole doping ($Ca^{2+}$ on $La^{3+}$ site). The increase in superconducting transition temperature with increasing dopant concentration confirms that the two phenomenon (hole filling and hole doping) are non-compensatory and hole doping dominates over hole filling [5]. The structural analysis has shown that in all LPCaBCO compounds, Ca is distributed at both La and Ba site and Pr is fully occupying La site only with no observed displacement onto Ba site. The ionic differences between different cations at La and Ba site can give rise to lattice strain resulting in an increase in the inhomogeneity of superconductivity and enhancement of critical current density ($J_c$).

In this paper we report the results of magnetization and susceptibility measurements performed on all samples of LPCaBCO series. The observed increase in critical current density ($J_c$) and flux pinning ($F_p$) are discussed in context of increase in pinning centers created by the additional holes contributed to the system with increasing $Ca^{2+}$ concentration. These additional holes play an important role in enhancing superconducting transition temperature and also in increasing the flux pinning and $J_c$.

**Experimental**

All the samples of LPCaBCO series were prepared by solid-state reaction method using high purity starting compounds (purity > 99.99%). Reacting constituents were taken in proportionate quantities and ground thoroughly under acetone. Samples were calcined twice at $930^0C$ for 24 hours each. The resultant powders were subjected to thorough grinding and pelletization. The pellets were sintered in atmosphere at $940^0C$ for 24 hours followed by first annealing in flowing $O_2$ at $940^0C$ for 12 hours, and at $500^0C$ for 24 hours with slow cooling at the rate of $1^0C$/min till room temperature. The samples were characterized by X-ray diffraction for single-phase formation. The structural investigations were carried out by X-ray Rietveld refinement method [5, 15]. Magnetization measurements were carried out on a commercially available SQUID (Quantum Design) magnetometer at TIFR, Mumbai.

**Results and Discussion**

The critical current densities were determined from the width of the hysteresis loops, measured at 5 K for all samples using the Bean's critical state model [16]. The critical current density ($J_c$ in $A/cm^2$) can be expressed in M (emu) and calculated to first approximation by the following relation [17],

$$J_c = \frac{30[M^+ - M^-]}{D} \quad ---(1)$$

$M^+$ and $M^-$ are the magnetization values observed during the up and down cycle

Correspond Author: Dr. S. Rayaprol (TIFR, Mumbai) sudhindra@tifr.res.in,
Phone Number: +91-22-2280 4545 (Ext: 2439)

of field sweeping, $D$ was estimated by SEM and assumed to be spherical grain, of size ~ $2.3 \times 10^{-4}$ cm.

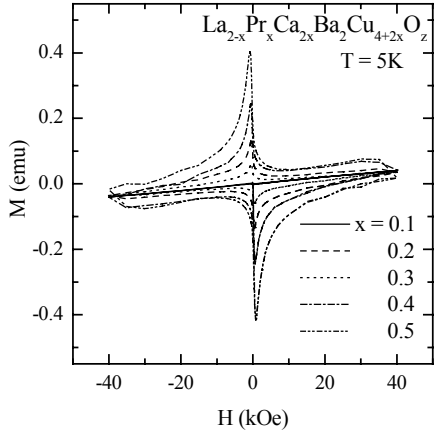

***Fig. 1*** *Hysteresis loops for $La_{2-x}Pr_xCa_{2x}Ba_2Cu_{4+2x}O_z$ samples with x = 0.1 – 0.5*

Figure 1 shows the hysteresis curves for all the samples recorded at 5 K. The decrease of $J_c$ as a function of field is shown in Figure 2.

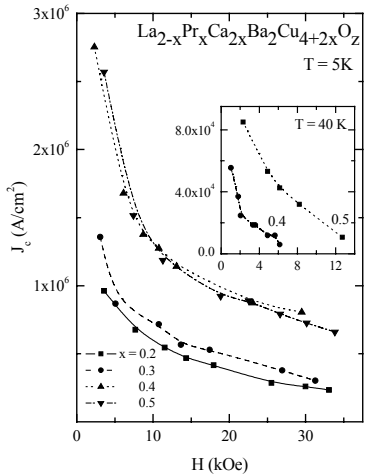

***Fig 2*** *Variation of current density ($J_c$) for $La_{2-x}Pr_xCa_{2x}Ba_2Cu_{4+2x}O_z$ samples with x = 0.2 – 0.5 under the applied field, measured at 5 K. The insert shows the current density measured at 40 K for x = 0.4 and 0.5 samples. The connecting lines are guides for eyes.*

It can also be seen that $J_c$ decreases with increasing temperature, while $J_c$ increases with increasing dopant concentration. The increase in $J_c$ is compared with the increase in $T_c$ as shown in Figure 3.

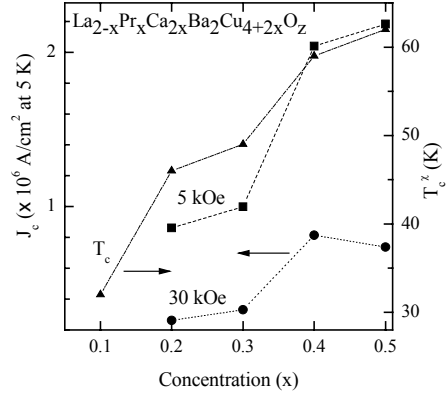

***Fig 3*** *Change in critical current density ($J_c$) as a function of dopant concentration (x) observed at 5 and 30 kOe. For comparison we have plotted the variation of $T_c$ (observed from onset of diamagnetic signal from dc susceptibility plot) with x.*

It is interesting to observe here that for low field (5 kOe), highest critical current density is observed for x = 0.4 sample, while at high field (30 kOe) the x = 0.5 sample exhibits little higher current density. The critical current density in the mixed state is related to the bulk flux pinning force ($F_p$) by

$$F_p = J_c \times B \quad \text{---(2)}$$

where B is the magnetic induction. $F_p$ can arise due to structural defects in the sample. Assuming that $J_c$ is perpendicular to B, we have calculated the bulk flux pinning force as

$$F_p = J_c B \quad \text{---(3)}$$

Using relation (3) we have calculated $F_p$ at 5 K for all the samples in the

Correspond Author: Dr. S. Rayaprol (TIFR, Mumbai) sudhindra@tifr.res.in,
Phone Number: +91-22-2280 4545 (Ext: 2439)

LPCaBCO series and plotted against H, in Figure 4.

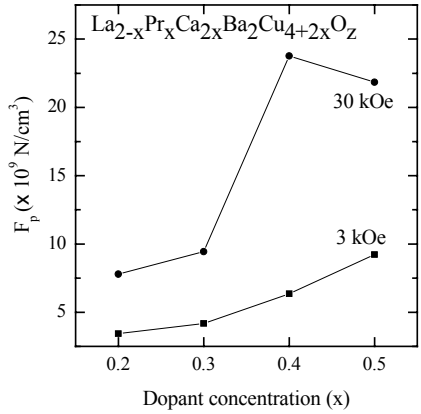

**Fig 4** *Change in critical current density ($F_p$) as a function of dopant concentration (x) observed at 3 and 30 kOe.*

The increase in pinning force with increasing dopant concentration and applied field can be clearly seen in Figure 5.

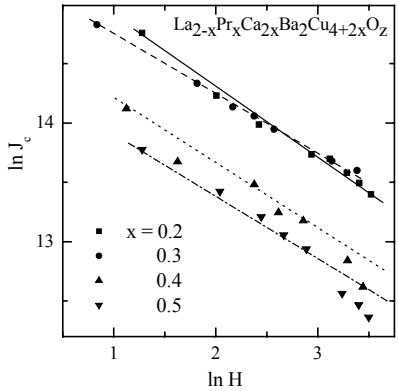

**Fig 5** *Slope of $lnJ_c$ vs. $lnH$ for all samples in $La_{2-x}Pr_xCa_{2x}Ba_2Cu_{4+2x}O_z$ series with x = 0.2 – 0.5. The plot satisfies $J_c \propto H^{0.5}$ relation, with the slopes (n) equal to 0.5 for all the compounds, thus exhibiting 'flux creep'. Straight lines are drawn as guides for eyes.*

The behavior of $J_c$ in magnetic fields can be explained by flux creep model [18]. The critical current density ($J_c$) follows a $H^{-1/2}$ behavior as predicted by

$$J_c(H,t) = \frac{N_p F_p(H,0)}{1.07(\phi_0 H)^{1/2}}\left[1 - \alpha(H)t - \beta t^2\right]$$

---(4)

where $\alpha$ and $\beta$ are coefficients, $N_p$ is the density of the pinning sites and $\phi_0$ is the flux quantum. In order to verify the validity of flux creep model for LPCaBCO systems, we have plotted $lnJ_c$ versus lnH in Figure 6. The slopes are in the acceptable range (~ 0.5) of flux creep model, for all the samples. The values of transition temperature (inferred for susceptibility plots), critical current density ($J_c$), flux pinning energy ($F_p$) and slope (n) from $lnJ_c$ vs. lnH plot are summarized in Table 1.

**Conclusion**

The superconducting transition temperature ($T_c$), critical current density ($J_c$) and flux pinning energy ($F_p$) increases with increasing dopant concentration in LPCaBCO system. We assume that the dopants play a dominant role in the creation of holes in the system inducing superconductivity. The holes created by these dopants acts as pinning centers, thus enhancing flux pinning and critical current density. Thus, it can be concluded that in La-2125 superconductors there exists a strong dependence of nature of dopants and their concentration on the enhancement in superconducting properties, which is reflected from the increase in $T_c$, $J_c$ and $F_p$ with dopant concentration in LPCaBCO system.

Correspond Author: Dr. S. Rayaprol (TIFR, Mumbai) sudhindra@tifr.res.in,
Phone Number: +91-22-2280 4545 (Ext: 2439)

Table 1   Values of critical current density ($J_c$), flux pinning energy ($F_p$), Superconducting transition temperature ($T_c^{onset}$) and slope (n) of $\ln J_c$ versus $\ln H$

| Concentration (x) | Critical current density ($J_c$ in $10^6$ A/cm$^2$) | | Flux pinning energy ($F_p$ in $10^9$ N/cm$^3$) | | Transition temperature ($T_c^{onset}$) K (± 1K) | Slope (*n*) |
|---|---|---|---|---|---|---|
| | 5 kOe | 30 kOe | 3 kOe | 30 kOe | | |
| 0.1 | -- | -- | -- | -- | 32 | -- |
| 0.2 | 0.86 | 0.26 | 3.44 | 7.79 | 46 | 0.52 |
| 0.3 | 0.99 | 0.33 | 4.18 | 9.44 | 49 | 0.52 |
| 0.4 | 2.03 | 0.81 | 6.35 | 23.76 | 59 | 0.50 |
| 0.5 | 2.18 | 0.73 | 9.22 | 21.85 | 62 | 0.57 |

**Acknowledgements**
SR and DGK thankfully acknowledge the financial assistance from IUC-DAEF (Mumbai) in the form of a collaborative research project (CRS-M-88). Authors also thank Prof. S. K. Malik and Dr. D. C. Kundaliya for magnetization measurements.

Correspond Author: Dr. S. Rayaprol (TIFR, Mumbai) sudhindra@tifr.res.in,
Phone Number: +91-22-2280 4545 (Ext: 2439)

Correspond Author: Dr. S. Rayaprol (TIFR, Mumbai) sudhindra@tifr.res.in,
Phone Number: +91-22-2280 4545 (Ext: 2439)